\documentclass[preprint,aps,tightenlines,showpacs,nofootinbib,superscriptaddress]{revtex4}
\usepackage{latexsym}
\usepackage[dvips]{graphicx}
\newcommand{\beq}{\begin{eqnarray}}
\newcommand{\eeq}{\end{eqnarray}}
\newcommand{\eq}{eqnarray}

\newcommand{\al}{{\alpha}}

\newcommand{\ci}{\cite}

\newcommand{\ep}{{\epsilon}}

\newcommand{\de}{{\delta}}
\newcommand{\De}{\Delta}

\newcommand{\Th}{{\Theta}}

\newcommand{\ka}{{\kappa}}
\newcommand{\la}{{\lambda}}
\newcommand{\La}{{\Lambda}}

\newcommand{\om}{{\omega}}
\newcommand{\Om}{{\Omega}}
\newcommand{\pa}{{\partial}}
\newcommand{\no}{{\nonumber}}
\newcommand{\f}{\frac}
\newcommand{\ra}{\rightarrow}

\newcommand{\Sch}{Schwarzschild }

\newcommand{\fn}{\footnote}

\newcommand{\calB}{{\cal B}}
\newcommand{\calC}{{\cal C}}

\newcommand{\calG}{{\cal G}}
\newcommand{\calH}{{\cal H}}

\newcommand{\calR}{{\cal R}}

\newcommand{\Ho}{Ho\v{r}ava~}
\newcommand{\cA}{{\cal A}}
\newcommand{\cB}{{\cal B}}
\newcommand{\cC}{{\cal C}}
\newcommand{\cE}{{\cal E}}

\newcommand{\cG}{{\cal G}}
\newcommand{\cH}{{\cal H}}

\newcommand{\cR}{{\cal R}}
\newcommand{\cS}{{\cal S}}

\begin{document}

\preprint{arXiv:1701.03844v2 [hep-th]}

\title{On Gauge Invariant Cosmological Perturbations
in UV-modified Ho\v{r}ava Gravity
}

\author{Sunyoung Shin}
\affiliation{208, Nokwon A. Sangga, 30, Dunsannam-ro,
Seo-gu, Daejeon, 35235, Korea}

\author{Mu-In Park\footnote{E-mail address: muinpark@gmail.com, Corresponding author}}
\affiliation{Research Institute for Basic Science, Sogang University,
Seoul, 121-742, Korea }

\begin{abstract}
We consider gauge invariant cosmological perturbations in UV-modified, $z=3$ (non-projectable) Ho\v{r}ava gravity with one scalar matter field,
which has been proposed as  a renormalizable gravity theory without the
ghost problem in four dimensions.
In order to exhibit its dynamical degrees of freedom, we consider the Hamiltonian reduction method and find that, by solving {\it all} the constraint equations, the degrees of freedom are the same as those of Einstein gravity: One scalar and two tensor (graviton) modes when a scalar matter field presents. However, we confirm that there is no extra graviton modes and general relativity is recovered in IR, which achieves the consistency of the model. From the UV-modification terms which break the detailed balance condition in UV, we obtain scale-invariant power spectrums for
{\it non}-inflationary backgrounds, like the power-law expansions,
without knowing the details of early expansion history of Universe. This could provide a new framework for the Big Bang cosmology. Moreover, we find that {\it tensor and scalar fluctuations travel differently in UV, generally.} We present also some clarifying remarks about confusing points in the literatures.
\end{abstract}


\maketitle
\newpage

\section{Introduction}

Several years ago, Ho\v{r}ava proposed a renormalizable, higher-derivative gravity theory, without ghost problems,
by considering different scaling dimensions for space and time
\ci{Hora}. From the lack of full diffemorphism ({\it Diff}) beyond general relativity (GR) limit, its constraint structure is quite complicated and not completely understood yet \ci{Li,Henn,Bell}.

As a closely related problem, there have been confusions regarding the extra graviton mode, known as the `scalar graviton', and the recovery of GR in IR limit \ci{Char,Blas:0906,Koya,Blas:0909,Papa,Moon}. However, in the non-projectable case, which allows arbitrary space-time dependent linear perturbations in the lapse function $N$ so that there exists the ``local" Hamiltonian constraint as well as the momentum constraints, it has been shown that the scalar mode can be consistently decoupled from the usual tensor graviton modes in the spatially flat ($k=0$) cosmological background as well as the flat (vanishing cosmological constant, $\La=0$) Minkowski vacuum for some appropriate gauge choices \ci{Gao,Keha,Park:0910,Kim}. Later, a more complete analysis has been done by considering the most general expressions of the cosmological perturbations for metric as well as that of a canonical scalar matter field and it has been shown that, without choosing any gauge, there are one scalar and two tensor modes as in GR, by solving {\it all} the constraint equations in the Hamiltonian reduction method \ci{Gong} \footnote{In the projectable case \ci{Gong}, where the lapse function is a function of time only, or in the extended model with the dynamical lapse function \ci{Blas:0909,Koba}, there exists one extra scalar graviton mode. But in this paper, we will not consider those cases  since it is not clear whether it could be a viable model even in our solar system, not to mention pathological ghost behaviors \ci{Koya,Papa}. }.

However, there is a serious physical problem in those works since
scale-invariant power spectrums for the quantized cosmic scalar fluctuations can not be obtained, in disagreement with observational data. This is basically due to the exact cancelation of the sixth-order-spatial derivative terms for scalar fluctuations, from the ``detailed balance" condition, which was originally introduced to construct the four-dimensional, power-counting renormalizable gravity action with some limited number of independent coupling constants, motivated by reminiscent methods in condensed matter systems \ci{Hora}.

On the other hand, it is already known that the detailed balance condition is too restrictive to get a viable model in IR limit since the usual \Sch black hole solution as well as the Newtonian potential in the weak field approximation for vanishing cosmological constant can not be obtained. Similar to the IR problem, one can also cure the above UV problem by considering  some ``UV-breaking" of the detailed balance condition so that the sixth-spatial derivatives for the scalar fluctuations do not cancel.

In this regard, there have been several works already for the flat Minkowski vacuum \ci{Park:0910,Soti,Bogd}, FRW cosmology \ci{Soti,Son,Muko:1007},  cosmological perturbations \ci{Qiu} ( \ci{Chen,Wang:0907} for the projectable case, \ci{Koh,Ceri,Cai} for the extended case), but still a through analysis, similar to \ci{Gong} is lacking. One of the purpose of this paper is to fill the gap by considering the gauge invariant cosmological perturbations in the Hamiltonian reduction approach, without the above mentioned UV problem.

In this paper, we consider gauge invariant cosmological perturbations in
UV-modified (non-projectable) Ho\v{r}ava gravity with one scalar matter field so that the
scalar cosmological fluctuation can be also (power-counting) renormalizable
with the dynamical critical exponent $z=3$ in four dimensions. By solving
{\it all} the constraints using the Hamiltonian reduction method, we find
that only one scalar and two tensor (graviton) degrees of freedom are left
when a scalar matter field presents, as in Einstein gravity. This confirms
that there is no extra scalar graviton mode and GR is recovered in IR, which
provides the consistency of the model. Furthermore, we obtain the
scale-invariant power spectrums for {\it non}-inflationary
backgrounds, like the power-law expansions with the power of $1/3<p<1$, without knowing the details of early expansion history of Universe. Moreover, we find that {\it tensor and scalar fluctuations travel differently in UV, generally.} This could provide a new framework for the Big Bang cosmology. In addition, we revisit several debating issues which have been discussed  earlier within the Lagrangian approach with the appropriate gauge choices in \ci{Gao} and confirm their resolutions in the Hamiltonian approach in a gauge independent way. We have also clarified several confusing points which have not been discussed clearly in \ci{Gong}.

\section{UV-modified \Ho action and its cosmological perturbations}

We start by considering the ADM decomposition of the metric,
\begin{\eq}
ds^2=-N^2 c^2 d\eta^2
+g_{ij}\left(dx^i+N^i d\eta\right)\left(dx^j+N^j d\eta\right)\,
\end{\eq}
and the UV-modified \Ho  gravity action with $z=3$, {\it a `la} Ho\v{r}ava,
which is power-counting renormalizable \ci{Hora}, is given by
\begin{\eq}
\label{HL action}
S_\mathrm{g} &=& \int d \eta d^3x \sqrt{g} N \left[ \frac{2}{\kappa^2}
\left(K_{ij}K^{ij} - \lambda K^2 \right) - {\cal V} \right], \\
-{\cal V}&=& \sigma+ \xi R + \alpha_1 R^2+ \alpha_2 R_{ij}R^{ij}
+\alpha_3 \frac{\epsilon^{ijk}}{\sqrt{g}}R_{il}\nabla_jR^l{}_k \no \\
 &+& \alpha_4 \nabla_{i}R_{jk} \nabla^{i}{R}^{jk}
+\al_5 \nabla_{i}R_{jk}\nabla^{j} {R}^{ik}
+\al_6 \nabla_{i}R\nabla^{i}R  
, \label{V_Horava}
\end{\eq}
where
\begin{\eq}
K_{ij}=\frac{1}{2N}\left({g_{ij}}'
-\nabla_i N_j-\nabla_jN_i\right)\
\end{\eq}
is the extrinsic curvature (the prime $(')$ denotes the derivative with
respect to $\eta$), $\ep^{ijk}$ is the Levi-Civita symbol, $R_{ij}$ and $R$
are the  Ricci tensor and scalar of the three-dimensional (Euclidean) spacial
geometry, respectively, and $\kappa,\lambda,\xi,\alpha_i$ are coupling constants
\footnote{The notations differ from those of \ci{Gong} as $\kappa^2=2
\kappa^2_\mathrm{(Here)}, \mu=\xi_\mathrm{(Here)},\al_1=\al_{2 \mathrm{(Here)}},
\al_2=\al_{1 \mathrm{(Here)}},\al_4=-\al_5=-8 \al_6=\al_{4 \mathrm{(Here)}}$.}.
From the prescription of the detailed balance condition, the number of
independent coupling constants can be reduced to six, {\it i.e.},
$\kappa,\lambda,\mu,\nu,\La_W,\om$ for a viable model in IR \ci{Keha,Park:0910},
\begin{\eq}
\label{detailed balance}
&&\sigma=\f{3 \kappa^2 \mu^2 \La_W^2}{8 (3 \la-1)},~
\xi=\f{\ka^2 \mu^2 (\om-\La_W)}{8 (3 \la-1)},~
\al_1=\f{\ka^2 \mu^2 (4 \la-1)}{32 (3 \la-1)},~
\al_2=-\f{\ka^2 \mu^2 }{8},~
\al_3=\f{\ka^2 \mu }{2 \nu^2},\no~\\
&&\al_4=-\f{\ka^2}{8 2 \nu^4}=-\al_5=-8 \al_6,
\end{\eq}
in contrast to four fundamental constants in GR, {\it i.e.}, Newton constant $G$,
speed of light $c$, cosmological constant $\La$, and the fixed coupling constant, $\la=1$.
However, in the below we do not restrict to this case only, at least for the
UV couplings $\al_4,\al_5,\al_6$ so that the power-counting renormalizable
and scale-invariant cosmological scalar fluctuations can be obtained.

For the power-counting renormalizable matter action, we consider $z=3$ scalar
field action \ci{Calc,Kiri},
\begin{equation}
\label{matter action}
S_\mathrm{m} = \int d\eta d^3x \sqrt{g} N \left[ \frac{1}{2N^2}
 \left( \phi' - N^i \pa_i \phi\right)^2 - V(\phi)- Z(\pa_i\phi)  \right] \, ,
\end{equation}
where
\begin{equation}
\label{Z action}
Z(\pa_i\phi) = \sum_{n=1}^3 \xi_n \partial_i^{(n)}\phi \partial^{i(n)}\phi \, ,
\end{equation}
with the superscript $(n)$ denoting $n$-th spatial derivatives, and
$V(\phi)$ is the matter's potential without derivatives.

The actions (\ref{HL action}) and (\ref{matter action}) are invariant under
the foliation preserving {\it Diff} \ci{Hora}
\begin{\eq}
\label{Diff}
\delta x^i &=&-\zeta^i (\eta, {\bf x}), ~\delta \eta=-f(\eta), \no \\
 \delta
g_{ij}&=&\pa_i\zeta^k g_{jk}+\pa_j \zeta^k g_{ik}+\zeta^k
\pa_k g_{ij}+f g'_{ij},\nonumber\\
\delta N_i &=& \pa_i \zeta^j N_j+\zeta^j \pa_j N_i+\zeta^{'j}
g_{ij}+f {N'}_i+f' N_i, \no \\
\delta N&=& \zeta^j \pa_j N+f N'+f' N, \no \\
\delta \phi&=&\zeta^j \pa_j \phi +f \phi'.
\end{\eq}

In order to study the cosmological perturbations around the homogeneous and isotropic backgrounds, we expand the metric and the scalar field as,
\begin{\eq}
\label{pert}
N &=&  a(\eta)[1+{\cal A}(\eta,{\bf x})] \, ,~
N_i =  a^2(\eta){{\cal B}(\eta,{\bf x})}_i \, ,~
g_{ij} =  a^2(\eta) [\de_{ij}+h_{ij} (\eta,{\bf x})],\\ \no
\phi &=& \phi_0(\eta) + \delta\phi(\eta,{\bf x}) \, ,
\end{\eq}
by considering spatially flat ($k=0$) backgrounds and the conformal (or {\it comoving}) time $\eta$, for simplicity \fn{For the metric with the physical time $dt=a d \eta$, $ds^2=-{\cal N}^2 c^2 d\eta^2 +g_{ij}(dx^i+{\cal N}^i d\eta)(dx^j+{\cal N}^j d\eta)$, one can obtain ${\cal N}=N/a=(1+\cA),~ {\cal N}_i=N_i/a=a \cB_i$.}. By substituting the metric and scalar
field of (\ref{pert}) into the actions one can obtain the linear-order
perturbation part of the total action $S=S_g +S_m$ as follows
(up to some boundary terms)
\begin{\eq}
\label{HL linear}
\delta_1 S
&=& \int d \eta d^3x~ a^2
\left\{
\left[ -\frac{6(1-3\lambda)}{\kappa^2}\calH^2- \left(\f{1}{2 } \phi_0'^2+ a^2(V_0
-\sigma) \right)\right] {\cal A} \right.\no \\
 &+&\left.\f{1}{2}\left[-\frac{2(1-3\lambda)}{\kappa^2} \left(\calH^2+2\calH'\right)
 +\left(\f{1}{2 }\phi_0'^2-a^2(V_0-\sigma) \right)\right] h \right.\no \\
 &-& \left. \left(\phi_0''+2 {\cal H} \phi_0' +a^2 V_{\phi_0'}\right) \de \phi
\right\} \, ,
\end{\eq}
which results the background equations, known as the Friedman's equations,
\begin{\eq}
\label{friedmann_eq}
 &&\calH^2 =
-\frac{\kappa^2}{6(1-3\lambda)} \left( \f{1}{2}{\phi_0'}^2
 + a^2 \left(V_0 -\sigma \right) \right) \, ,
\\
\label{H_evolution}
&&  \calH^2 +2\calH' =
\frac{\kappa^2}{2(1-3\lambda)} \left( \f{1}{2}{\phi_0'}^2
 - a^2 \left(V_0 -\sigma \right) \right) \, ,
\\
\label{phi_evolution}
&&\phi_0'' + 2\calH\phi_0' + a^2V_{\phi_0} = 0 \, ,
\end{\eq}
with the comoving Hubble parameter ${\cal H} \equiv a'/a$, $V_0 \equiv V(\phi_0), V_{\phi_0}\equiv (\pa V/\pa \phi)_{\phi_0},h\equiv h^i_i$, and indices raised and lowered by $\de_{ij}$.
Here, it important to note that there is {\it no} higher-derivative corrections to the background equations of (\ref{friedmann_eq}) and (\ref{H_evolution}) for spatially flat case and so the background equations are the same as those of GR \ci{Keha}. However, even in this case, the higher-derivative effects can appear in the perturbed parts. Moreover, we note that the spatial curvature $k$ and cosmological constant $\Lambda$ can be independent parameters only when either the space-time is time-dependent or matter exists, which has been confused sometimes in the literatures. \footnote{In \ci{Hora}, the metric perturbations were considered around (spatially flat and static) Minkowski vacuum ($k=0, \La=0$) but the vacuum solution can not be the solution of the gravity with detailed balance ! \ci{Lu,Keha,Park:0905}}

The quadratic part of the total perturbed action is given by
\begin{\eq}
&&\de_2 S = \int d \eta d^3x \left\{ \f{2 a^2}{\ka^2} \left[ (1-3 \la) {\cal H} \left(
3 {\cal H} {\cal A}^2 + {\cal A} (2 \pa {\cal B}^i-h') \right)
+ (1-\la) (\pa_i {\cal B}^i)^2+\f{1}{2} \pa_i {\cal B}_j \pa^i {\cal B}^j \right. \right. \no \\
&&~~\left. \left. -\pa_i {\cal B}_j {h^{ij}}' +\f{1}{4} {h_{ij}}' {h^{ij}}'
+\la \left(\pa_i {\cal B}^i h'-\f{1}{4}h'^2 \right) \right]
+a^2 \xi \left({\cal A} +\f{1}{2} h \right)
\left( \pa_i \pa_j h^{ij} -\Delta  h \right) \right.  \no \\
&&~~\left.+a^2 \left[ \f{1}{2} \de \phi'^2-{\cal A} \phi_0' \de \phi' +\f{1}{2} {\cal A}^2 \phi_0'^2 +\pa_i {\cal B}^i \phi_0' \de \phi-\f{a^2}{2} V_{\phi_0 \phi_0} \de\phi^2 -a^2 V_{\phi_0} \de \phi {\cal A}-\f{1}{2} \phi_0' \de \phi h'\right] \right.\no \\
&&~~\left. -a^4 \left({\cal V}^{(2)}+\de Z \right)\right\},
\end{\eq}
where $\Delta  \equiv \de^{ij} \pa_i \pa_j$ is the spatial Laplacian, $\de Z=\sum_{n=1}^3 \xi_n~ \partial_i^{(n)}\de\phi \partial^{i(n)}\de\phi$, and ${\cal V}^{(2)}$ is the quadratic part of the potential ${\cal V}$ in (\ref{V_Horava}).

Now, in order to separate the scalar, vector, and tensor contributions, we consider the most general decompositions,
\begin{\eq}
\label{decomposition_0i}
\calB_i &=&  \pa_i {\cal B} + S_i \, ,
\no \\
\label{decomposition_ij}
h_{ij} &=& 2 {\cal R} \de_{ij}+ \pa_i \pa_j {\cal E} + \pa_{(i} F_{j)} + \tilde{H}_{ij} \, ,
\end{\eq}
where $S_i$ and $F_i$ are transverse vectors,
and $\tilde{H}_{ij}$ is a transverse-traceless tensor, {\it i.e.,}
\begin{equation}
\pa_i S^i = \pa_i F^i = \tilde{H} = \pa_i \tilde{H}^i_j = 0 \, .
\end{equation}
Then, the pure tensor, vector, and scalar parts of the total action is given by, respectively,
\begin{\eq}
\label{reduced_2nd_tensor_action}
\delta_2 S^{(t)} &=& \int d \eta d^3x~ a^2
 \left[ \frac{2}{\kappa^2} {\tilde{H}_{ij}}' {\tilde H}^{ij'}
+\xi\tilde{H}_{ij}\Delta \tilde{H}^{ij}
+ \frac{\alpha_2}{a^2}\Delta \tilde{H}_{ij}\Delta \tilde{H}^{ij}+ \frac{\alpha_3}{a^3} \epsilon^{ijk} \Delta \tilde{H}_{il} \Delta  \pa_j \tilde{H}^l_{k}  \right. \no \\
&&~~~~~~~~~~~~~~~~~~~~\left.
 - \frac{\alpha_4}{a^4} \Delta \tilde{H}_{ij}\Delta^2 \tilde{H}^{ij} \right],\\
\label{reduced_2nd_vector_action}
\delta_2 S^{(v)} &=& \frac{1}{\kappa^2} \int  d \eta d^3x~ a^2 \pa_i
 \left( S^j-{F^j}' \right) \pa_i \left( S_j-F_j' \right) \, ,\\
\label{reduced_2nd_scalar_action}
\delta_2 S^{(s)}
&=&  \int d \eta d^3x ~a^2 \left\{ \frac{2(1-3\lambda)}{\kappa^2}
 \left[ 3{\calR'}^2 - 6\calH {\cal A}\calR' +
3\calH^2 {\cal A}^2
- 2\left( \calR' - \calH \cA \right) \Delta({\cal B}-{\cal E}')
\right] \right.
\nonumber\\
 &&+ \frac{2(1-\lambda)}{\kappa^2} \left[ \Delta\left({\cal B}-{\cal E}'\right) \right]^2
- 2\xi(\calR+2{\cal A})\Delta\calR
+ \frac{2}{a^2} \left( 8\alpha_1+3\alpha_2  \right)(\Delta\calR)^2
\nonumber\\
&& \left.
-\frac{2}{a^4} \left( 3\alpha_4+2\alpha_5+8 \al_6 \right)\Delta\calR \Delta^2 \calR -a^2V_{\phi_0} \cA\delta\phi
  -\frac{1}{2a^2}V_{\phi_0\phi_0}\delta\phi^2 - \delta{Z} \right.\no \\
&& \left.+ \frac{1}{2}{\delta\phi'}^2  - \phi_0'\delta\phi' {\cal A}
+\f{1}{2} {\phi_0'}^2 {\cal A}^2
+ [\Delta({\cal B}-{\cal E}')-3 \calR']\phi_0'\delta\phi \right\} \, .
\end{\eq}
Here, it is important to note that sixth-order-derivative terms in the gravity action (\ref{HL action}), which are required in the power-counting renormalizability in four dimensions, contribute to scalar as well as tensor perturbations, through the specific combination of `$3\alpha_4+2\alpha_5+8 \al_6$' for the former but through only  `$\al_4$' for the latter. This implies that
\\
\\
$~~~$ {\it tensor and scalar perturbations travel differently in UV, generally.}
\\
\\
On the other hand, unlike the tensor and scalar parts, the vector perturbation is not dynamical, from the lack of kinetic terms.

\section{Hamiltonian reduction and dynamical degrees of freedom}

In order to exhibit the true dynamical degrees of freedom we consider the Hamiltonian reduction method \ci{Fadd}, for the cosmologically perturbed actions (\ref{reduced_2nd_tensor_action})-(\ref{reduced_2nd_scalar_action}) \ci{Garr,Gong}.

For the tensor part, the reduction is rather trivial since it is already in the unconstrained form with two physical modes which can be interpreted as two polarizations of the (primordial) gravitational waves as in GR \ci{Keha,Park:0910,Star,Taka,Koh:0907}. In other words, the perturbed action (\ref{reduced_2nd_tensor_action}) can be written as the first-order form,
\begin{\eq}
&&\delta_2 S^{(t)} = \int d \eta d^3x
\left( \Pi^{ij} {\tilde{H}_{ij}}' - \calH^{(t)} \right)\,,
\\
&&\calH^{(t)}=
\frac{\kappa^2}{8a^2}\Pi^{ij}\Pi_{ij}
- a^2\xi \tilde{H}^{ij}\Delta \tilde{H}_{ij}
- \alpha_2\Delta \tilde{H}^{ij}\Delta \tilde{H}_{ij}
-\frac{\alpha_3}{a} \epsilon^{ijk}\Delta \tilde{H}_{il}\Delta \pa_j \tilde{H}^l{}_{k} \no \\
&&~~~~~~~~ +\frac{\alpha_4}{a^2}\Delta \tilde{H}^{ij}\Delta^2\tilde{H}_{ij}\, ,
\label{2nd_tensor_action:Hform}
\end{\eq}
with the conjugate momentum,
\begin{equation}
\label{momentum_hij}
\Pi^{ij} \equiv \frac{\delta (\delta_2 S^{(t)})}{\delta {\tilde{H}_{ij}}'}
=\frac{4 a^2}{\kappa^2} {\tilde{H}}^{ij'} \, ,
\end{equation}
but without any constraint terms.

Next, for the vector part, the action (\ref{reduced_2nd_vector_action}) can written as the first-order form,
\begin{\eq}
\label{vec_action_1st_order}
\delta_2 S^{(v)} &=&  \int d\eta d^3x \left( \Pi^i \Delta F_i'
-\calH^{(v)} \right)\,,
\\
\calH^{(v)}&=&-\frac{\kappa^2}{4 a^2}\Pi^i \Delta \Pi_i +\Pi^i \Delta  S_i\, ,
\label{vec_Hamil_1st_order}
\end{\eq}
with the conjugate momentum,
\begin{equation}
\Pi^i \equiv \frac{\delta (\delta_2 S^{(v)}) }{\delta \Delta F_i'}
 = \frac{2 a^2}{\kappa^2} \left( S^i-{F^i}' \right) \, .
\end{equation}
But, from the equations of motion for $\Delta  S_i$,
\begin{equation}
0=-\f{\de \calH^{(v)}}{\de \Delta  S_i}= \Pi^i  \, ,
\label{vectorconstraints}
\end{equation}
which is a constraint equation, one obtains the vanishing action (\ref{vec_action_1st_order}) and its Hamiltonian (\ref{vec_Hamil_1st_order}), which show the non-dynamical nature of the vector perturbation.

Finally, for the scalar part, which is the most non-trivial one, the action (\ref{reduced_2nd_scalar_action}) can be written  as the first-order form, after some computations,
\begin{\eq}
\label{1st_order_L}
\de_2 S^{(s)} &=& \int d\eta d^3x
\left( \Pi_{\delta\phi}\delta\phi'
+\Pi_\calR \calR' + \Pi_{\Delta {\cal E}} {\Delta {\cal E}}'- \calH^{(s)}
-{\cal A}\calC_{\cal A} - \Pi_{\Delta {\cal E}} \Delta {\cal B}  \right) \, ,
\\
\calH^{(s)} &=&  \frac{\kappa^2}{8a^2}
 \left[-\Pi_\calR \Pi_{\Delta {\cal E}}
 + \frac{3}{2} \Pi_{\Delta {\cal E}}^2 +
\frac{4}{\kappa^2}{\Pi_{\delta\phi}}^2
 + \frac{1-\lambda}{2(1-3\lambda)}{\Pi_\calR}^2\right]
 + \frac{\kappa^2}{4(1-3\lambda)}\phi_0' \Pi_\calR \delta\phi
\nonumber\\
&&
+ \frac{3\kappa^2 a^2}{8(1-3\lambda)}{\phi_0'}^2\delta\phi^2
+ a^2 \left( \delta{Z} + \f{a^2}{2} V_{\phi_0 \phi_0}\delta\phi^2 \right)
+ 2 a^2\xi\calR\Delta\calR
-2(8\alpha_1+3\alpha_2)(\Delta\calR)^2 \no \\
&&+\frac{2}{a^2} \left( 3\alpha_4+2\alpha_5+8 \al_6 \right)
\Delta\calR \Delta^2 \calR ,
\\
\calC_{\cal A} &=& \Pi_{\delta\phi}\phi_0' + \calH\Pi_\calR +  4a^2\xi\Delta\calR
+ a^2 \left(3\calH\phi_0' + a^2V_{\phi_0} \right)\delta\phi \, ,
\label{C_A}
\end{\eq}
with the conjugate momenta
\begin{\eq}
\label{Pi_deltaphi1}
\Pi_{\delta\phi} &\equiv& \frac{\delta (\delta_2 S^{(s)}) }{\delta \de \phi'}
=  a^2 \left( \delta\phi' - \phi_0' {\cal A} \right) \, ,
\\
\label{Pi_psi1}
\Pi_\calR &\equiv& \frac{\delta (\delta_2 S^{(s)}) }{\delta \calR'}
=  a^2 \left\{ \frac{4(1-3\lambda)}{\kappa^2} \left[ 3\left(
\calR'-\calH {\cal A} \right) - \Delta({\cal B}-{\cal E}') \right]
-3\phi_0'\delta\phi \right\} \, ,
\\
\label{Pi_E1}
\Pi_{\Delta {\cal E}} &\equiv& \frac{\delta (\delta_2 S^{(s)} )}{\delta  \Delta {\cal E}'}
=  a^2  \left\{ \frac{4(1-3\lambda)}{\kappa^2}( \calR'-\calH {\cal A})
-\frac{4(1-\lambda)}{\kappa^2}\Delta({\cal B}-{\cal E}')
-\phi_0'\delta\phi \right\} \, .
\end{\eq}
Here, we note that ${\cal A}^2$ terms in (\ref{reduced_2nd_scalar_action}) are canceled  {\it without} using the background equation (\ref{friedmann_eq}) \footnote{This is in contrast to the on-shell result in \ci{Gong}.} and only the linearly-dependent terms remain in (\ref{1st_order_L})  so that ${\cal A}$, as well as ${\cal B}$, be the Lagrange multiplier.

Then, from the equations of motion for ${\cal A}$ and $\Delta {\cal B}$, which give the following constraint equations, respectively,
\begin{\eq}
{\cal C_A}=0,~~ \Pi_{\Delta {\cal E}}=0,
\end{\eq}
one obtains the reduced action, by eliminating $\Pi_\calR$ in (\ref{1st_order_L}),
\begin{\eq}
\label{1st_order_action:reduced 1}
&&\de_2 S^{(s)}_\mathrm{red}~[\de \phi, \Pi_{\de \phi}; {\cal R}] = \int d\eta d^3x
\left( \Pi_{\delta\phi}\delta\phi'- \calH^{(s)}_\mathrm{red} \right) \, ,
\\
&&\calH^{(s)}_\mathrm{red} =  \frac{1}{2a^2}
 \left(1+ \frac{\kappa^2 (1-\la)}{8(1-3\lambda)}\f{{\phi_0'}^2}{{\cal H}^2} \right){\Pi_{\delta\phi}}^2
 + \frac{\kappa^2 (1-\la)}{8(1-3\lambda)}\phi_0' (3 \phi_0' {\cal H}+a^2 V_{\phi_0} )~\delta\phi \Pi_\calR
\nonumber\\
&&~~~+ \f{a^4}{2} \left[ \frac{\kappa^2 (1-\la)}{8(1-3\lambda)}\left(3 {\phi_0'}+\f{a^2 V_{\phi_0}}{{\cal H}} \right)^2+
\frac{3 \kappa^2 }{4(1-3\lambda)}{\phi_0'}^2 + V_{\phi_0 \phi_0} \right]\delta\phi^2 + a^4 \delta{Z} \no \\
&&~~~+ \frac{\kappa^2 (1-\la)}{2(1-3\lambda)} \xi\phi_0' \Pi_{\delta\phi} \Delta {\cal R}
+ \frac{a^2 (1-\la)}{2(1-3\lambda)} \xi (3 \phi_0' {\cal H}+a^2 V_{\phi_0} )~\delta\phi \Delta \calR
+ 2 a^2\xi\calR\Delta\calR \no \\
&&~~~+\left[\frac{\kappa^2 (1-\la)}{(1-3\lambda)}\f{a^2 \xi^2}{{\cal H}^2}  -2(8\alpha_1+3\alpha_2)\right](\Delta\calR)^2
+\frac{2}{a^2} \left( 3\alpha_4+2\alpha_5+8 \al_6 \right)
\Delta\calR \Delta^2 \calR ,
\end{\eq}
which shows one dynamical scalar degree of freedom for the matter $\de \phi$ and one non-dynamical field  ${\cal R}$ from the metric.

In order to obtain a gauge-invariant description for the true dynamical degrees of freedom, we consider a new variable,
\begin{\eq}
\label{zeta}
\zeta \equiv  \delta\phi- \frac{\phi_0'}{\calH}\calR \, ,
\end{\eq}
which is gauge invariant since ${\cal R}$ transforms as $\de \cR=\cH f$ under the foliation preserving {\it Diff} (\ref{Diff}), and its conjugate momentum,
\begin{\eq}
\label{Pi_eta}
\Pi_\zeta \equiv
\Pi^{\delta\phi} - \frac{a^3}{\calH}\left( \frac{\phi_0'}{a} \right)'\calR .
\end{\eq}
Then, after some analysis, we find that the reduced action (\ref{1st_order_action:reduced 1}) can be written as
\begin{\eq}
&&\delta_2 \hat{S}_\mathrm{red}^{(s)}~[\zeta, \Pi_{\zeta}; {\cal R}]
= \int d\eta d^3x
\left[ \Pi_{\zeta}\zeta'- \hat{\calH}^{(s)}_\mathrm{red} (\zeta, \Pi_{\zeta}; {\cal R}) \right] \, ,
\\
&&\hat{\calH}_\mathrm{red}^{(s)}(\zeta, \Pi_{\zeta}; {\cal R}) =
A(\zeta, \Pi_{\zeta}) - B(\zeta, \Pi_{\zeta})\Delta\calR + \Delta\calR \Theta \calR  \, ,
\label{Ht:red}
\end{\eq}
where
\begin{eqnarray}
A(\zeta, \Pi_{\zeta}) &= & A_1 {\Pi_{\zeta}}^2 + A_2 \Pi_{\zeta} \zeta + \zeta A_3 \zeta \, ,
\label{A}\\
B(\zeta, \Pi_{\zeta}) &= & B_1 \Pi_{\zeta} + B_2 \zeta\, ,
\label{B}\\
\Theta(\zeta, \Pi_{\zeta}) &=& \Theta_1+\Theta_2 \Delta + \Theta_3 \Delta^2,
\label{Theta}
\end{eqnarray}
whose explicit forms are given in Appendix A.

From the equations of motion for $\De \cR$,
\begin{equation}
\label{eqR}
0=\f{\de \cH^{(s)}}{\de \De \cR}=2 \Theta \calR-B \equiv \cC_{\De \cR},
\end{equation}
one can eliminate $\cR$ also and finally obtain the physical action, with only the physical variables $\zeta$ and $\Pi_{\zeta}$,
\begin{\eq}
&&\delta_2 {S}_\star^{(s)}~[\zeta, \Pi_{\zeta}]
= \int d\eta d^3x
\left[ \Pi_{\zeta}\zeta'- {\calH}_{\star}^{(s)} (\zeta, \Pi_{\zeta}) \right] \, ,
\label{phys action}\\
&&{\calH}_\star^{(s)}(\zeta, \Pi_{\zeta}) =
A(\zeta, \Pi_{\zeta}) - \f{1}{4} B(\zeta, \Pi_{\zeta})\Delta \Theta ^{-1} B(\zeta, \Pi_{\zeta})   \, ,
\label{phys H}
\end{\eq}
where the {\it non-local} operator $\Theta^{-1}$ is defined as $\Th\Th^{-1}=1$, which may satisfy $\Th^{-1}\Th=1$ when the zero mode can be ignored. This completes the Hamiltonian reduction of the system, which ends up with only the gauge invariant, physical degrees of freedom.

The usual second-order action form is given by
\begin{equation}
\label{phys action2}
\delta_{2}S_\star^{(s)}
 = \int d \eta d^3x \left\{\f{1}{4} \zeta'\calG_{1}^{-1}\zeta'
 + \zeta\left[ \f{1}{4}\left( \calG_{2}{\calG_1}^{-1} \right)'
 + \f{1}{4}(\calG_2)^2{\calG_1}^{-1} - \calG_3 \right]\zeta \right\} \, ,
\end{equation}
where
\begin{equation}
\calG_1 \equiv A_1 - \f{1}{4}(B_1)^2\Delta \Th^{-1} \, ,
\quad
\calG_2 \equiv A_2 - \f{1}{2}B_1 B_2 \Delta \Th^{-1} \, ,
\quad
\calG_3 \equiv A_3 - \f{1}{4}(B_3)^2\Delta \Th^{-1} \, ,
\end{equation}
using the Hamilton's equation,
\begin{\eq}
\zeta'=\{ \zeta, \int d^3x \cH^{(s)}_{\star} \}=2 \cG_1 \Pi_{\zeta} + \cG_2 \zeta,
\end{\eq}
with the Poisson bracket,
\begin{\eq}
\{ \zeta(\eta, {\bf x}), \Pi_{\zeta}(\eta, {\bf y}) \}=\de ^3 ({\bf x}-{\bf y}),
\end{\eq}
which can be read from the symplectic structure in the action (\ref{phys action2}) \ci{Witt}.

By introducing a new canonical variable,
\begin{equation}
u \equiv (2\calG_1)^{-1/2} \zeta \, ,
\end{equation}
the action (\ref{phys action2}) can be written as
\begin{equation}
\label{uaction}
\delta_2S_\star^{(s)}
 = \int d\eta d^3x \frac{1}{2}\left\{ {u'}^2
 - u \left[ \f{1}{2}\left( \calG_1'\calG_1^{-1} \right)'
 - \f{1}{4}\left(\calG_1'\calG_1^{-1}\right)^2
 - \calG_1 \left( \calG_2 \calG_1^{-1} \right)'
 - \calG_2^2 + 4\calG_1\calG_3 \right]u \right\} \, .
\end{equation}
with its equations of motion as
\begin{\eq}
\label{uequation}
&&u'' =- \om_u^2 u  \, , \\
&&\om_u^2=  \f{1}{2}\left( \calG_1'\calG_1^{-1} \right)'
 - \f{1}{4}\left(\calG_1'\calG_1^{-1}\right)^2
 - \calG_1 \left( \calG_2 \calG_1^{-1} \right)'
 - \calG_2^2 + 4\calG_1\calG_3  .
\end{\eq}

\section{Comparison with observational data}

In order to be compared with some observational data about the early Universe, let us consider UV limit \fn{For IR limit, the usual Mukhanov equation $u''-\De u-(z''/z)u$ is obtained when we set $\ka^2 \mu^2=1$ and $\la=1$ \ci{Gong}.} of our cosmological perturbations. First, for the scalar perturbations, the field equation of the canonical scalar field $u$ (\ref{uequation}) reduces to, in UV limit\fn{The result (\ref{omega_UV}) can be also checked in the Lagrangian approach of \ci{Gao}: For the UV-modified terms in the potential part of (\ref{V_Horava}), the corrected coefficients in UV limit are $\tilde{d}_{\Psi}=g_3 \dot{\varphi}^2/H^2-6 \tilde{\al}_4, \Om=-(g_3 \dot{\varphi}^2/H^2)[1-(1-6 \tilde{\al}_4 H^2/g_3 \dot{\varphi}^2)^{-1}] \De^3$ and then one can obtain the same UV limit as in (\ref{omega_UV}) \ci{Moon:priv}.},
\begin{\eq}
\label{uequation_UV}
&&u'' =- \om_{u(UV)}^2 u  \, , \\
\label{omega_UV}
&&\om_{u(UV)}^2=  \f{-6 \xi_3 \tilde{\al_4}}{a^2 z^2 } \left[ 2+\f{\ka^2 (1-\la)}{4 (1-3 \la)}\f{z^2}{a^2}\right] \De^3,
\end{\eq}
where $3 \tilde{\al_4} \equiv 3 \al_4 +2 \al_5+8 \al_6, ~z \equiv a \phi_0'/\cH$. Here, it is important to note that there are sixth-spatial derivatives, as required by the scale invariance of the observed power spectrum \ci{Muko:1007,Gesh} as well as the (power-counting) renormalizability \ci{Hora}. This occurs only when there are sixth-derivative terms in the starting scalar action (\ref{matter action}), (\ref{Z action}) ({\it i.e.,} $\xi_3 \neq 0$ in (\ref{Z action})) as well as some breaking of the detailed balance condition in sixth-derivative terms for the gravity action (\ref{HL action}) ({\it i.e.,} $\tilde{\al_4} \neq 0$) \fn{The UV limit of (\ref{omega_UV}) differs from that of the scalar graviton, $\om^2 \sim (1-\la) \tilde{\al}_4 \De^3/(1-3 \la)$, which vanishes in the GR limit of $\la \ra 1$ \ci{Park:0910,Chen}.}.

Regarding the scale invariance of the power spectrums, it has been noted that  Ho\v{r}ava gravity could provide an alternative mechanism for the early Universe without introducing the hypothetical inflationary epoch \ci{Muko:1007,Kiri,Muko:0904}: The basic reason of the alternative mechanism comes from the momentum-dependent speeds of gravitational perturbations which could be much larger than the current, low energy ({\it i.e., }IR) speed $c$ so that the exponentially expanding early spacetime could be mimicked. \fn{This idea is reminiscent of, so-called, the ``varying speed of light (VSL)" model \ci{Moff,Albr}. This fact seems to be another justification for the direction of Ho\v{r}ava gravity, other than the original motivation for renormalizability \ci{Hora}.} Especially, it has been argued that even the power-law expansions \ci{Lucc} ($t$ is the physical time, defined by $dt=a d \eta$),
\begin{\eq}
a =a_0 t^p, ~(1/3 < p <1)
\end{\eq}
could produce the scale-invariant power spectrums \cite{Muko:0904,Kiri}. (For an earlier discussion, see also \ci{Calc}). In this case, from the relations,
\begin{\eq}
a&=&a_0^{1/(1-p)} [(1-p) \eta]^{p/(1-p)},
\label{a by eta}
\\
z^2&=&\f{(1-3 \la)}{\ka^2} \f{a^2}{\cH^2} (\cH'-\cH^2) \no \\
&=&-\f{(1-3 \la)}{\ka^2} \f{a_0^{2/(1-p)}}{p} { [(1-p) \eta]^{2p/(1-p)}},
\label{z by eta}
\end{\eq}
[ we have used the background equation (\ref{H_evolution}) in the first line of (\ref{z by eta}) ] one finds that the UV frequency in (\ref{omega_UV}) can be written as
\begin{\eq}
\label{omega_UV_power}
\om_{u(UV)}^2=  \f{-6 \xi_3 \tilde{\al_4} \ka^{2}}{(1-3 \la) a^4 } \left( 2p-\f{1-\la}{4 }\right) \De^3
\end{\eq}
with $\xi_3 \tilde{\al_4}[ 2p-(1-\la)/4]/(3 \la-1)\geq 0$ for a stable 
perturbation.

Moreover, for some analytic computations, we consider only one interesting case of $p=1/2$, which corresponds to the radiation-dominated era at the early Universe. Then, it is easy to see that the normalized mode function is given by \fn{This is the only case where $a''/a$ and $z''/z$ vanish.}\ci{Muko:1007,Star}
\begin{\eq}
u_k &=&\sqrt{\f{\hbar{\cal M}^2}{2 k^3}}~ a~ \mathrm{exp} \left(-\f{ik^3}{{\cal M}^2} \int^\eta \f{d \eta}{a^2} \right), \\
{\cal M}^2 &=&\sqrt{\f{-(1-3 \la)}{6 \ka^2 \xi_3 \tilde{\al}_4 [1-(1-\la)/4]}}~, \no
\end{\eq}
from the standard normalization condition,
\begin{\eq}
\left< u_k, u_k\right> \equiv \f{i}{\hbar} (u^*_k u'_k-{u^*_k}'u_k)=1
\end{\eq}
with the Fourier expansion,
\begin{\eq}
u(\eta, {\bf x}) =\int \f{d {\bf k}^3}{(2 \pi)^3} u_{\bf k} (\eta) e^{i {\bf k} \cdot {\bf x}}.
\end{\eq}

Then, the (dimensionless) power spectrum \fn{For some introductory materials about the power spectrum, see \ci{Baum}, for example.} $\De_{\zeta}^2 (k)$ for the quantum field $\hat{\zeta}$ of the $\zeta$ perturbation,
\begin{\eq}
\left<0|
\hat{\zeta}_{\bf k} (\eta) \hat{\zeta}_{{\bf k}'} (\eta)
|0\right>
=(2 \pi)^3 \de ({\bf k}+{\bf k}') \f{2 \pi^2}{k^3} \De^2_{\zeta} (k)
\end{\eq}
is obtained as
\begin{\eq}
\De^2_{\zeta}&=&\f{k^3}{2 \pi^2} | \zeta_{\bf k} |^2
\no \\
&=&\f{\hbar}{8 \pi^2} \sqrt{\f{(3 \la-1)[1-{(1-\la)}/{4}] }{6 \ka^2 \xi_3 \tilde{\al}_4}}~,
\end{\eq}
using
\begin{\eq}
\zeta^2&=&2 {\cal G}_1 u^2 \no \\
&\approx& \f{1}{2 a^2} \left[1-\f{(1-\la)}{4} \right] u^2
\end{\eq}
in UV limit, without knowing the details of the history of the early
Universe and the form of the (non-derivative) potential $V (\phi)$
\fn{The only assumption is the existence of the renormalizable $z=3$
(Lifshitz) scalar field as well as the Ho\v{r}ava-Lifshitz's gravity field.
The Higgs field, which is the only currently known fundamental scalar field,
would be a strong candidate for the scalar field and it would be an important question whether
the Higgs field can be the late-time remnant of the primordial scalar field
on the expanding Universe, in our new context of cosmology.}. Note that
there is no $k$ ({\it i.e.}, scale) dependence in the UV limit but
its $k$-dependence appears in the subleading terms, which produce the
running of the spectral indices. By identifying the UV power spectrum with
the observed (nearly) scale-invariant scalar power spectrum,
$\De^2_{\zeta}
\sim 10^{-9}$,
one finds that
\begin{\eq}
\label{alpha_4_tilde}
\xi_3 \tilde{\al}_4 \sim  \f{10^{13} \hbar^2 (3 \la-1)}{\ka^2} \left[1-\f{(1-\la)}{4} \right].
\end{\eq}

Now, for the tensor perturbation, the similar result can be obtained more easily. To do this, we first note that the UV limit of the field equation for transverse traceless mode $\tilde{H}_{ij}$ is given by
\begin{\eq}
\label{Hequation_UV}
&&\tilde{H}_{ij}'' =- \om_{H(UV)}^2 \tilde{H}_{ij}  \, , \\
\label{H_omega_UV}
&&\om_{H(UV)}^2=  \f{\ka^2 \al_4}{2 a^4} \De^3,
\end{\eq}
with $\al_4\geq 0$ for the stable tensor perturbation, whose propagation speed is different from that of the scalar fluctuation in (\ref{uequation_UV}) and (\ref{omega_UV}) generally, as anticipated in the action forms of (\ref{reduced_2nd_tensor_action}) and (\ref{reduced_2nd_scalar_action}).
Then, it is easy to find that the {\it canonically} normalized mode function is given by
\begin{\eq}
\sqrt{\f{8}{\ka^2}}h^s_k &=&\sqrt{\f{\hbar \tilde{\cal M}^2}{2 k^3}}~ \mathrm{exp} \left(-\f{ik^3}{\tilde{\cal M}^2} \int^\eta \f{d \eta}{a^2} \right), \\
\tilde{\cal M}^2 &=&\sqrt{\f{2}{ \ka^2 \al_4}}, \no
\end{\eq}
with the Fourier expansion,
\begin{\eq}
H_{ij}(\eta, {\bf x}) =\int \f{d {\bf k}^3}{(2 \pi)^3}\sum_{s=+,\times} \ep_{ij}^s h^s_{\bf k} (\eta) e^{i {\bf k} \cdot {\bf x}}.
\end{\eq}
Then, one can obtain the total power spectrum for the perturbations (the factor ``2" is due to the two polarizations of the tensor perturbation, $s=+,\times$),
\begin{\eq}
\De^2_t&=&2 \De^2_h \no \\
&=&\sum_{s=+,\times}\f{k^3}{2 \pi^2} | h^s_{\bf k} |^2 \no \\
&=&\f{\hbar \ka^2}{8 \pi^2 \sqrt{2 \ka^2 \al_4}},
\end{\eq}
which is also scale-invariant, with the polarization tensor $\ep_{ij}^s$ satisfying $\ep^s_{ii}=k^i \ep^s_{ij}=0$ and $\ep^s_{ij}(k) \ep^{s'}_{ij}(k)=2 \de_{ss'}$.

The tensor-to-scalar ratio is obtained as
\begin{\eq}
r &\equiv& \f{\De^2_t}{\De^2_\zeta} \no \\
&=&\ka^2 \sqrt{\f{ (3 \xi_3 \tilde{\al}_4/\al_4)}{(3\la-1)[1-(1-\la)/4]}}.
\end{\eq}
On the other hand, from the known values of $\De^2_{\zeta}$, $\al_4$ can also be expressed as
\begin{\eq}
\al_4 \sim \f{10^{-4} \hbar^2 \ka^2}{\De^4_t}
 \sim \f{10^{14} \hbar^2 \ka^2}{r^2 }.
\label{alpha_4}
\end{\eq}
So, for the large values of the tensor-to-scalar ratio, $r \ge 0.01$, one can obtain
\begin{\eq}
\label{alpha_4_b}
\al_4 \stackrel{<}{\sim} 10^{18}\hbar^2 \ka^2.
\end{\eq}
From the intimate relation between the UV conformal symmetry in the
Ho\v{r}ava gravity with the detailed balance condition and the particular
coupling of $\la=1/3$ \ci{Hora}, some quantum mechanical breaking of the UV
conformal symmetry, {\it i.e., conformal anomaly}, could be one possible
origin of the breaking of the UV detailed balance condition. If this is
the case, one can obtain $ka^4 \xi_4 \gg 10^{-5}$ by assuming that other
UV parameters $\al_5$ and $\al_6$ are also the same order as $\al_4$,
{\it i.e.}, $ \stackrel{<}{\sim} {10^{18}}\hbar^2 \ka^2$, but their
linear combination $3 \tilde{\al_4} \equiv 3 \al_4 +2 \al_5+8 \al_6\sim 10^{13}
\hbar^2 (3 \la -1) [1-(1-\la)/4]/\ka^2 \xi_3$ is relatively tiny
! Moreover, even though one can not estimate about RG-flows for these parameters separately since all the coupling parameters $\al_i,\ka,\la$, and $\xi_n$ could flow under RG, the results of (\ref{alpha_4_tilde}), (\ref{alpha_4}), and (\ref{alpha_4_b}) in UV limit, contrast to the IR limit of ($c \equiv 1$), $\al_i \ra 0, \xi_1 \ra 1/2, \xi_{2,3}\ra 0, \la \ra 1, \ka^2 \ra 32 \pi G, \xi \ra 1$ for the appropriate IR-modified action \ci{Argu}, seems to indicate the {\it real} occurrence of RG-flows.

Now, in order to understand how the scale-invariant UV fluctuations are related to the current cosmological observables, we first note that, in our case of $1/3 < p < 1$, the conformal (or comoving) time is given by, from (\ref{a by eta}),
\begin{\eq}
\eta=\left[(1-p) a_0^{(1-p)/p}\right]^{-1} a^{(1-p)/p}
\end{\eq}
and one can take the initial moment as $\eta_i=0$ with $a(\eta_i)=0$, in contrast to the $p>1$ power-law expansion or the usual inflation theory with the exponential expansion. Here, note that, due to the momentum-dependent, superluminal speeds of fluctuations, from (\ref{omega_UV}) and (\ref{omega_UV_power}), one does not need to introduce the infinite past as the initial moment, {\it i.e.}, $\eta_i=-\infty$ in order to resolve the casual communication problem in the early Universe, like the ``horizon problem" in the Big Bang cosmology: The past light cones in the early Universe, which is assumed to be UV region, are more flattened than that of GR in IR region so that any two points at the surface of the observed CMB, $\eta_{CMB} \sim 0.03 \eta_0$ for the current time $\eta_0$, can be in causal contact at the initial moment of the Big Bang, $\eta_i=0$ (Fig. 1).
\begin{figure}
\includegraphics[width=12cm,keepaspectratio]{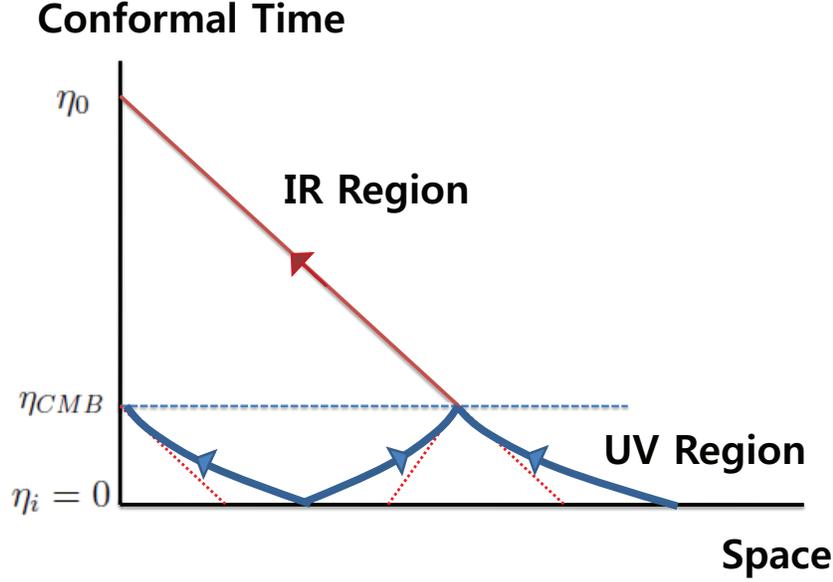}
\caption{Conformal diagram of the Big Bang Universe. Two points at the observed CMB $\eta_{CMB}$, which may be belong to IR region, can be in causal contact at the Big Bang, $\eta_i=0$, due to flattened past light-cones from the superluminal speeds of fluctuations in UV region (blue lines). The dotted red lines denote the usual light cones with the same speeds of fluctuations as in IR and show the casually disconnected regions at the Big Bang.}
\label{fig:E:eta}
\end{figure}

Next, we note that, while we have considered $\om_{UV} \gg \cH$ in UV limit, $\om_{UV}$ decreases slowly with $\om_{UV}'/\om_{UV}=-4 \cH \ll \om_{UV}$ so that the fluctuations follow adiabatically the initially scale-invariant oscillations until $\om_{UV} \sim k^3/a^2 \sim \cH$ is reached, where the adiabaticity breaks down and the fluctuations start to freeze-out. This transition point corresponds to the horizon exit in the conventional inflation theory (Fig. 2). After some evolution in the freezing region outside the horizon in UV region, the fluctuations re-enter the horizon in IR region, where $\om_{IR} \sim k$, since $\cH \sim a^{-(1-p)/p}\sim \om_{IR}$ for any power-law expansion with $1/3<p<1$ so that the scale-invariant fluctuations with the usual IR dispersion can be observed now \ci{Muko:1007}.

This can be also understood in the comoving length scales, where the relevant comoving Hubble length is given by $v_{UV} \cH^{-1}\sim k^2 a^{-2} \cH^{-1} \sim k^2 a^{(1-3p)/p}$ and $v_{IR} \cH^{-1}=\cH^{-1}\sim k^2 a^{(1-p)/p}$ with the characteristic velocities $v_{UV}$ and $v_{IR}$, for the UV and IR regions, respectively, with a fixed comoving length scale $k^{-1}$ (Fig. 3). In this picture, it is clear why we need the condition `$1/3<p<1$' for the viable Big Bang cosmology model with the horizon exit and re-entry: The relevant Hubble length at UV, $v_{UV} \cH^{-1}$ is {\it decreasing} due to the decreasing characteristic velocity $v_{UV}\sim k^2 a^{-2}$ during expansions for $1/3<p$ so that there is a horizon exit, whereas {\it increasing} IR Hubble length $v_{IR} \cH^{-1}=\cH^{-1}$ for $p<1$ so that there is ``necessarily" a horizon re-entry. Note that the conditions of the horizon exit and re-entry agree with those of corresponding energy scales, as described above and Fig. 2 \footnote{One can also consider the physical length scale analysis with $\om^{\mathrm{phy}}=\om/a,k^{\mathrm{phy}}=k/a, H\equiv\dot{a}/a=\cH/a$ and then one finds the relevant physical Hubble length as $v^{\mathrm{phy}}_{UV} H^{-1}\sim k^2 a^{-2} H^{-1} \sim k^2 a^{(1-2p)/p}$ and $v^{\mathrm{phy}}_{IR} H^{-1}=H^{-1}\sim k^2 a^{1/p}$ for UV and IR regions, respectively. Fig. 1 of \ci{Muko:1007} gives essentially the same result since the horizon crossing points, $k^3 a^{(1-3p)/p}\sim 1$, are exactly the same as ours, though the approach looks different from ours in UV. However, the comoving picture is not quite straightforward in that approach.}. It is remarkable that we could mimic the conventional inflationary picture without introducing the hypothetical inflationary epoch and other subsequent processes.

\begin{figure}
\includegraphics[width=12cm,keepaspectratio]{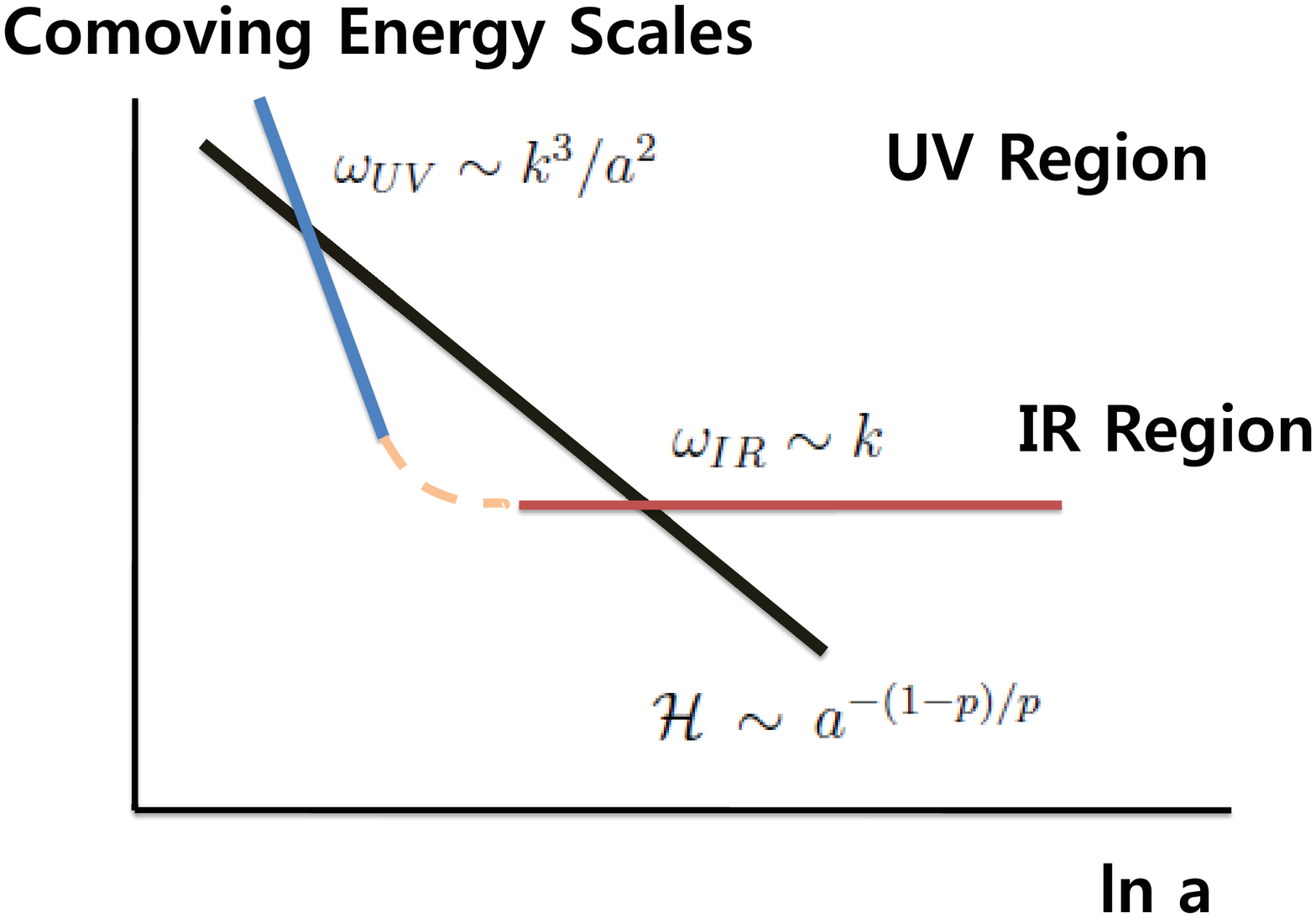}
\caption{Comoving energy scales $\om$ for fluctuations vs. the Hubble expansion rate $\cH$. For the power-law expansions with $1/3<p<1$, $\cH \sim a^{-(1-p)/p}$, there are two transition points where $\om$ meet $\cH$ which divide two (UV and IR) oscillating regions and one overdamping (or freezing) region in between. The UV and IR transition points correspond to the horizon exit and re-entry, respectively, in the conventional inflation theory.}
\label{fig:E:eta}
\end{figure}

\begin{figure}
\includegraphics[width=12cm,keepaspectratio]{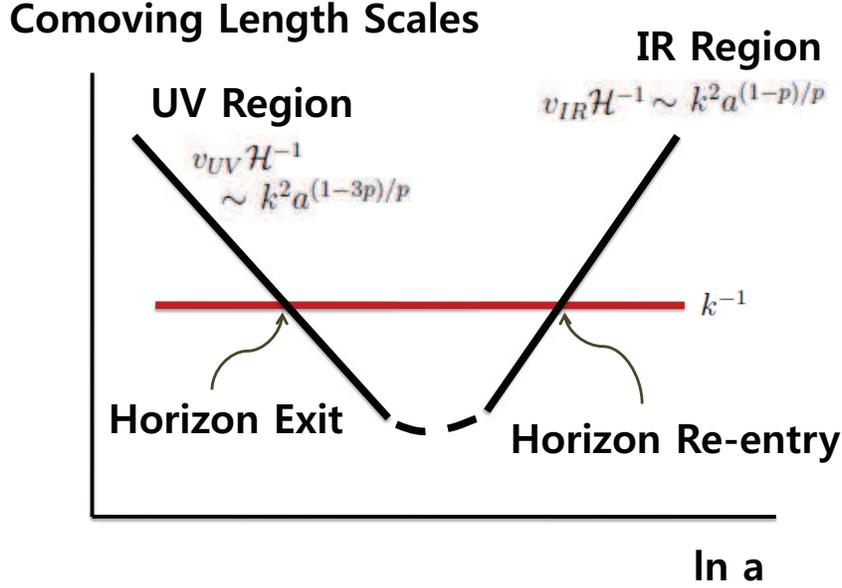}
\caption{Comoving length scales $k^{-1}$ for fluctuations vs. the relevant Hubble length (or horizon) scale $v\cH^{-1}$ with the characteristic velocity $v$. For the power-law expansions $a\sim \eta^{p/(1-p)}$ with $1/3<p$, there always exists a horizon exit at $v_{UV}\cH^{-1}\sim k^{-1}$ due to the decreasing characteristic velocity during expansions $v_{UV}\sim k^2 a^{-2}$ in UV region. Once there is the horizon exit, there is always the corresponding re-entry in IR region, due to ordinary IR Hubble length scale $v_{IR} \cH^{-1}=\cH^{-1}\sim k^2 a^{(1-p)/p}$ for $p<1$. These results agree with those of energy scales in Fig. 2. }
\label{fig:E:eta}
\end{figure}

\section{Concluding Remarks}

In conclusion, we have shown that only one scalar and two tensor (graviton) physical degrees of freedom in the cosmological perturbations for the UV-modified (non-projectable) Ho\v{r}ava gravity with a single scalar matter field, by solving {\it all} the constraints from the Hamiltonian reduction method. We have also shown that, without knowing the details of early expansion history of Universe, the power spectrums are scale invariant, which are in agreement with observational data, when there are the violation of UV detailed balance condition in the (power-counting) renormalizable $z=3$ Ho\v{r}ava gravity with one scalar matter field. This could provide a new framework for the Big Bang cosmology.

Now, several clarifying remarks are in order.

1. {\it On the constraint algebras and counting the number of degrees of
freedom}: For constrained systems, one can consider either the Dirac's
method without explicit solving the constraints or the Hamiltonian
reduction method when the constraints can be explicitly solved, or
generally some combinations of these two. In the Dirac's method, one
can obtain the reduced phase space through the Dirac bracketing with
classification of constraints as first and second-class, primary
constraints or their descendants, called secondary constraints \ci{Dira}.
On the other hand, in the Hamiltonian reduction method, the constrained
variables are eliminated by solving the constraints, {\it when applies},
so that only un-constrained variables remain in the reduced systems \ci{Fadd}.
In the later method, it is an important fact that the distinction between the
first-class constraints and the second-class constraints is not necessary since the algebras between
constraints are not needed. It is known that these two approaches are
equivalent ``classically" (see \ci{Garc}, for example), though quantum mechanically non-trivial. But the later approach is
more efficient in identifying the unconstrained variables, when applies
as in our case which will be discussed below in more details, without
resorting the cumbersome classifications of the first or second class. Moreover,
in the perturbed systems, we need to be more careful about the constraint
algebras since the more higher orders (at least one-order higher) than the
primarily obtained perturbative constraints from a given order of the
action or Hamiltonian. For example, in our case of quadratic-order systems,
one can obtain the first-order constraints, neglecting those of vector part
which is trivial, as
\begin{\eq}
\cC_{\cA}=0,~\cC_{\De \cB}\equiv \Pi_{\De \cE}=0,~\cC_{\De \cR}=0
\end{\eq}
but their constraint algebras produce the zero-th order quantities
through Poisson brackets. However, these are not enough to see whether the
algebras are closed ({\it i.e.}, producing already existing first-order constraints)
or not ({\it i.e.}, having some residual terms which can not be expressed by
the existing constraints), and this is the source of troubles in the constraint analysis
of \ci{Gong}.\footnote{This applies also to \ci{Park:0910} and \ci{Koh}.}
Actually, in \ci{Gong}, it is claimed that the constraints $\cC_{\cA}=0$ and
$\cC_{\De \cR}=0$ are the second-class constraints \footnote{$\cC_{\De \cR}=0$ corresponds
to the secondary constraint $\cC_2=0$ in \ci{Gong}, after the constraint
$\cC_{\cA}=0$ is used.} and $\cC_{\De \cB}=0$ is the first-class constraint but
this is a premature statement unless we get the first-order quantities through
Poisson brackets, which is not possible in the starting quadratic-order
systems.

However, since counting the number of degrees of freedom  through the
standard formula,
\begin{\eq}
s=\f{1}{2} (2n-2 N_1-N_2),
\end{\eq}
where $2n$ is the number of canonical variables, $N_1$ is the number of the
first-class constraints, and $N_2$ is the number of the second-class
constraints, produces $s=(2 \times 3-2 \times1-1 \times 3)/2=1$, in
agreement with the Hamiltonian reduction method in this paper, either the
naive estimation of \ci{Gong} could be confirmed even when considering the
higher-order effects, or there is one more ({\it i.e.}, fourth) constraint
which constitutes another second-class constraint with the used-to-be
first-class constraint, $\cC_{\De \cB}=0$ so that $N_1=0, N_2=4$. It would
be a challenging problem to investigate the constraint algebras by
considering cubic-order constraints.\\

2. {\it On the well-defindness of GR limit}: In the general decomposition of the metric, we start with four scalar degrees of freedom $\cA, \cB, \cE$, and $\cR$. The first three of them, $\cA, \cB, \cE$ are the Lagrange multipliers which are eliminated by solving or imposing their associated constraints $\cC_{\cA}=0,~\cC_{\De \cB}\equiv \Pi_{\De \cE}=0$. The fourth of them $\cR$ is eliminated by solving the constraint (\ref{eqR}). In this way, all the scalar degrees of freedom in the metric are eliminated ($\cA, \cB, \cE$ cases) or absorbed ($\cR$ case) into the gauge-invariant scalar variable $\zeta\equiv  \delta\phi- ({\phi_0'}/{\calH})\calR$, which corresponds to Mukhanov-Sasaki variable \ci{Mukh,Sasa}, so that the gauge transformation of the matter field $\de \phi$ is compensated by $\cR$. It is remarkable that {\it all} the constraints can be easily solved in the general cosmological backgrounds with $\cH, \phi_0' \neq 0$, which is quite unusual in the general context of gravity theory constraints.

On the other hand, even though the scalar part $\cR$ of the metric contributes to the gauge invariant variable $\zeta$, it is generated essentially by the scalar matter field through the coupling $({\phi_0'}/{\calH})$; in other words, the quantum perturbations of the scalar degrees of freedom are essentially coming from scalar matter fields as in the standard cosmological perturbation theory in GR and this confirms the similar statement in the Lagrangian approach with some appropriate gauge choices \ci{Gao}.

However, it is interesting to note that the {\it static} limit $\cH \ra 0$, {\it i.e.,}, flat Minkowski vacuum limit is {\it not} well-defined unless
${\phi_0'}\ra 0, \la \ra 1$ by looking at the physical Hamiltonian $\cH^{(s)}_{\star}$ in (\ref{phys H}) or its corresponding action $\cS^{(s)}_{\star}$ in (\ref{phys action2}). In particular, it is important to note that the purely GR limit of $\la \ra 1$ can be well-defined for arbitrary time-dependent backgrounds with $\cH \neq 0$: If one considers (incorrect) perturbations of a gravity system with matters around flat Minkowski vacuum, there would be singularities which signal a wrong choice of the Minkowski vacuum {\it when matters present} \ci{Char}.

As the last remark, the constraint $\cC_{\cA}=0$ in (\ref{C_A}), which corresponds to the ``local" Hamiltonian constraint, is essential to recover GR, as we have seen above (also in \ci{Park:0910}), and this is from allowing the arbitrary, space-time dependent fluctuations of the lapse function $N = a(\eta)(1+{\cal A}(\eta,{\bf x}))$. This is in contrast to the, so called, projectable case with $N = a(\eta)(1+{\cal A}(\eta))$, where there is no local Hamiltonian constraint but only its integrated form $\int d^3x \cC_{\cA}=0$ and recovering GR is not quite clear \ci{Chen}, as in the extended model \ci{Koba,Koh,Ceri,Cai}.

\appendix

\section{Explicit forms of $A,B,\Th$ coefficients in (\ref{A})-(\ref{Theta})}
\label{appendix_A}

In this appendix, we present the explicit forms of $A_i, B_j$, and $\Theta_k$ which appear in (\ref{A})-(\ref{Theta}):

\begin{\eq}
A_1 &=&  \frac{\kappa^2(1-\lambda)}{16a^2(1-3\lambda)}
  \frac{z^2}{a^4} + \frac{1}{2a^2} \, ,
\\
A_2 &=&  \frac{\kappa^2(1-\lambda)}{16(1-3\lambda)}
 \frac{z}{\calH}
\left( \frac{z \cH}{a^2} \right)' - \frac{\kappa^2}{4(1-3\lambda)}
 \left(  \frac{z^2 \cH}{a^2} \right) \, ,
\\
A_3 &=&  \frac{\kappa^2(1-\lambda)}{16(1-3\lambda)}
 \left[ \frac{a^2}{\calH} \left(  \frac{z \cH}{a^2} \right)' \right]^2
 + \frac{\kappa^2}{4(1-3\lambda)}a^2 z \left(  \frac{z \cH}{a^2} \right)'
\nonumber\\
&&~+ \frac{a^2}{2} \left[ \frac{3\kappa^2}{4(1-3\lambda)}{\phi_0'}^2
 + a^2V_{\phi_0\phi_0} \right] - a^2 \left(  \xi_1\Delta  - \xi_2\Delta^2
+\xi_3\Delta^3 \right) \, ,
\\
B_1 &=& -\frac{\kappa^2\xi(1-\lambda)}{2(1-3\lambda)}
\frac{z}{a \calH} \, ,
\\
B_2 &= & \frac{\kappa^2\xi (1-\lambda)}{(1-3\lambda)}
\frac{a^3}{\calH^2} \left(  \frac{z \cH}{a^2} \right)'
+ \frac{\kappa^2\xi}{1-3\lambda} a z
+ 2a z
\left(  \xi_1- \xi_2\Delta + \xi_3\Delta^2\right) \, ,
\\
\Th_1 &= & -z^2
\left( \xi_1 + \frac{\kappa^2\xi}{2(1-3\lambda)} \right) \, ,
\\
\Th_2 &= & z^2 \xi_2
+ \frac{\kappa^2\xi^2(1-\lambda)}{(1-3\lambda)}\f{a^2}{\calH^2}
- 2(8\alpha_1+3\alpha_2) \, ,
\\
\Th_3 &= & -z^2 \xi_3 -\f{2}{a^2} (3\alpha_4+2\alpha_5+8 \al_6)\, .
\end{\eq}
Here, we denote
\begin{\eq}
z \equiv \f{a \phi_0'}{\cH}=\sqrt{2 \ep} a,
\end{\eq}
with the slow-roll parameter $\ep$.

\section*{Acknowledgments}
MIP would like to thank Taeyoon Moon
for early collaboration and checking the UV limit of fluctuations in the Lagrangian approach \ci{Moon:priv}. This was supported by Basic Science Research Program through the National Research Foundation of Korea (NRF) funded by the Ministry of Education, Science and Technology {(2016R1A2B401304)}.

\newcommand{\J}[4]{#1 {\bf #2} #3 (#4)}
\newcommand{\andJ}[3]{{\bf #1} (#2) #3}
\newcommand{\AP}{Ann. Phys. (N.Y.)}
\newcommand{\MPL}{Mod. Phys. Lett.}
\newcommand{\NP}{Nucl. Phys.}
\newcommand{\PL}{Phys. Lett.}
\newcommand{\PR}{Phys. Rev. D}
\newcommand{\PRL}{Phys. Rev. Lett.}
\newcommand{\PTP}{Prog. Theor. Phys.}
\newcommand{\hep}[1]{ hep-th/{#1}}
\newcommand{\hepp}[1]{ hep-ph/{#1}}
\newcommand{\hepg}[1]{ gr-qc/{#1}}
\newcommand{\bi}{ \bibitem}

\end{document}